\documentclass[pra,twocolumn,showpacs,superscriptaddress,showkeys]{revtex4-1}

\usepackage{graphicx}
\usepackage{amsmath}
\usepackage{mathrsfs}
\usepackage{siunitx}
\usepackage{pstricks}
\usepackage{psfrag}
\usepackage{xspace}
\usepackage{color}
\usepackage{upgreek}
\usepackage[english]{babel}
\usepackage[normalem]{ulem}

\newcommand{\NoAutoSpaceBeforeFDP}{}
\newcommand{\AutoSpaceBeforeFDP}{}

\newcommand{\ie}{i.e.\@\xspace}

\newcommand{\eq}[1]{Eq.~\eqref{#1}}
\newcommand{\eqs}[1]{Eqs.~\eqref{#1}}

\newcommand{\fig}[1]{Fig.~\ref{#1}}
\newcommand{\figs}[1]{Figs.~\ref{#1}}
\newcommand{\Fig}[1]{Figure~\ref{#1}}

\renewcommand{\bm}[1]{\boldsymbol{\mathbf{#1}}}
\newcommand{\ud}{\mathrm{d}}
\newcommand{\bra}{\left\langle}
\newcommand{\ket}{\right\rangle}
\newcommand{\keff}{k_{\text{eff}}}
\newcommand{\im}{\operatorname{Im}}
\newcommand{\re}{\operatorname{Re}}

\newcounter{tempa}
\newcounter{tempb}
\newcounter{tempc}
\newcounter{tempd}

\newenvironment{diagc}[1]{\psset{unit=1.5mm,fillstyle=solid,fillcolor=white}
   \begin{pspicture}[shift=0.0](0,-1)(#1,2)}{\end{pspicture}
}
\newcommand{\ginc}[2]{\psline(#1,0)(#2,0)}

\newcommand{\particule}[1]{\pscircle(#1,0){1}}

\newenvironment{ddiag}[1]{\psset{unit=1.5mm,fillstyle=solid,fillcolor=white}
   \begin{pspicture}[shift=-5](0,-6)(#1,6)}{\end{pspicture}
}

\newcommand{\ggmoy}[3]{\psline[linewidth=0.5](#1,#3)(#2,#3)}
\newcommand{\eemoy}[3]{\psline[linewidth=0.5,linestyle=dashed](#1,#3)(#2,#3)}

\newcommand{\gggmoy}[4]{\psline[linewidth=0.5](#1,#3)(#2,#4)}

\newcommand{\pparticule}[2]{\pscircle(#1,#2){1}}
\newcommand{\nnonlineaire}[2]{
   \setcounter{tempa}{#1}
   \addtocounter{tempa}{-1}
   \setcounter{tempb}{#2}
   \addtocounter{tempb}{-1}
   \setcounter{tempc}{#1}
   \addtocounter{tempc}{1}
   \setcounter{tempd}{#2}
   \addtocounter{tempd}{1}
   \psframe(\value{tempa},\value{tempb})(\value{tempc},\value{tempd})
}
\newcommand{\iidentique}[4]{
   \psline(#1,#2)(#3,#4)
}

\newenvironment{dddiag}[1]{\psset{unit=1.5mm,fillstyle=solid,fillcolor=white}
   \begin{pspicture}[shift=-2](0,-3)(#1,15)}{\end{pspicture}
}
\newenvironment{ddddiag}[1]{\psset{unit=1.5mm,fillstyle=solid,fillcolor=white}
   \begin{pspicture}[shift=-17](0,-18)(#1,6)}{\end{pspicture}
}

\newenvironment{dddddiag}[1]{\psset{unit=1.5mm,fillstyle=solid,fillcolor=white}
   \begin{pspicture}[shift=-17](0,-18)(#1,18)}{\end{pspicture}
}

\begin{document}

\title{Photon echoes in strongly scattering media: a diagrammatic approach}

\author{R. Pierrat} 
\affiliation{ESPCI Paris, Universit\'e PSL, CNRS, Institut Langevin, 1 rue Jussieu, F-75005, Paris, France}
\email{romain.pierrat@espci.fr}
\author{R. Carminati}
\affiliation{ESPCI Paris, Universit\'e PSL, CNRS, Institut Langevin, 1 rue Jussieu, F-75005, Paris, France}
\author{J.-L. Le Gou\"et}
\affiliation{Laboratoire Aim\'e Cotton, CNRS, Universit\'e Paris-Sud, ENS Cachan, Universit\'e Paris-Saclay, F-91405, Orsay Cedex, France}

\begin{abstract}
    We study photon echo generation in disordered media with the help of multiple scattering theory based on
    diagrammatic approach and numerical simulations. We show that a strong correlation exists between the driving fields
    at the origin of the echo and the echo beam. Opening the way to a better understanding of non-linear wave
    propagation in complex materials, this work supports recent experimental results with applications to the
    measurement of the optical dipole lifetime $T_2$ in powders.
\end{abstract}

\maketitle 

\section{Introduction}\label{sec:introduction}

Wave scattering in disordered media has attracted considerable attention for decades. First undertaken within the
general scope of the multiple scattering theory, in close connection with quantum
mechanics~\cite{foldy1945,lax1951,lax1952}, investigations later focused on classical waves and optical processes,
revealing features such as the backscattering peak~\cite{albada1985,Wolf1985} or random lasing~\cite[and references
therein]{wiersma2008physics}.  Wave propagation in complex media can also be combined with nonlinear
optics~\cite{kravtsov1991theory,baudrier2004random}. In the specific framework of four-wave mixing (FWM), it has been
recognized quite early that coherent anti-Stokes Raman scattering (CARS) can take place in polycrystalline and opaque
media~\cite{Dlott1991}. The observation of wave localization, whether reduced, enhanced or simply tested by nonlinear
processes, definitely opens new
perspectives~\cite{deBoer1993,vanneste2001selective,cao2000spatial,wellens2008,wellens2009}.  

The temporal dimension is generally absent from these works. Indeed the scattered light emerges from the sample in close
synchrony with the incoming field, either because one operates far from the absorption lines, or because the lifetime of
the material resonances does not exceed the driving pulse duration, such as in CARS. The signature of the investigated
signal is obtained either in the angular scattering pattern or in the emission spectrum, the latter applying to
non-degenerate wave-mixing processes. 

Instead, we now consider a nonlinear signal that emerges from the sample long after the extinction of the driving
pulses. That time-domain discrimination may prove helpful in situations where neither the direction nor the wavelength
would enable to select the relevant scattered emission.  This time-delayed signal is produced by photon
echo~\cite{Kurnit1964,Abella1966}, a nonlinear process that belongs to the same four-wave mixing (FWM) class as
CARS~\cite{mossberg1982time}. Photon echo results from the resonant excitation of an absorbing line. The available time
delay is only limited by the optical dipole lifetime $T_2$ and may outdo the driving pulse duration by orders of
magnitude.  

Routinely used for $T_2$ measurement, a photon echo experiment is usually performed in samples of high optical quality.
However there is considerable practical interest to substitute a cheap and easily produced rough powder to a high
quality mono-crystal since interesting chemical solids are often difficult to
crystallize~\cite{markushev1990stimulated}. Such a simplified access to $T_2$ may expedite new compound selection in the
prospect of classical and quantum processing~\cite{colice2004rf,LE_GOUET-2007,tittel-photon,riedmatten2008}.   

The experimental observation of photon echo in rare earth ion doped polycrystalline powders at liquid helium
temperature~\cite{beaudoux2011}, and the successful demonstration of new compound testing~\cite{perrot2013}, call for a
better understanding of the scattered signal generation in such unusual conditions. In these studies, the echo is
efficiently detected by heterodyne mixing with a replica of one driving field. Hence, quite unexpectedly, two distinct
fields are able to preserve a strong correlation after erratic propagation through a disordered material although
the corresponding speckle patterns look very different.  The origin of such a disturbing and non intuitive
feature must be clarified. The present paper extends the well-established linear multiple scattering theory to the
non-linear, photon echo process. Special attention is paid to explaining the strong correlation of the echo with the
driving field.

The manuscript is organized as follows: in Sec.~\ref{sec:photon-echo} we summarize the main characteristics of photon
echoes. In Sec.~\ref{sec:model} we consider the case of a strongly disordered powder and we derive a physical model to
take into account photon echoes in such a material. In Sec.~\ref{sec:driving_theory} theoretical expressions for the
average driving fields and intensities are obtained. The theory is then expanded for the echo signal (average field,
average intensity and correlation with a driving field) in Sec.~\ref{sec:echo_theory}. Then we compare the analytical
results with numerical simulations in Sec.~\ref{sec:numerics} before concluding in Sec.~\ref{sec:conclusion}.

\section{Photon echo features}\label{sec:photon-echo}

Photon echo~\cite{Kurnit1964,Abella1966} refers to the time-delayed nonlinear coherent optical response to resonant
excitation by a specific sequence of light pulses.

In absorbing materials, $T_2$, the optical dipole lifetime, may be much larger than the inverse absorption bandwidth.
Indeed that bandwidth may reflect the Doppler shift, in gases, or a non-uniform transition frequency shift, caused by
interaction with the environment, in condensed matter, rather than the homogeneous linewidth. This quasi-static effect
is known as inhomogeneous broadening. When resonantly excited by a light pulse much shorter than $T_2$, the optical
dipole radiates a free induction decay (FID) signal. However this emission rapidly fades out because of inhomogeneous
phase shift, although optical dipoles keep on oscillating in the medium. The photon echo process, closely related to
spin echo in Nuclear Magnetic Resonance (NMR), is used to cancel the inhomogeneous phase shift and to recover a
radiative signature of the surviving dipoles.

Let us focus on stimulated echoes~\cite{mossberg1979}, generated by a sequence of three successive pulses that
resonantly excite an ensemble of two-level atoms. The pulses are labeled~1, 2 and 3, according to their time order. By
reducing the level population difference, resonant excitation partially bleaches the material over the pulse bandwidth.
However, bleaching caused by time-separated pulses is not uniform over the excitation bandwidth. In the same way
as, in space domain, two \emph{angled} beams can imprint a diffraction grating on a photographic plate, a pair of
\emph{time separated} pulses spectrally modulates the level population difference. Hence, pulses~1 and 2, separated by
time interval $t_{12}$, modulate the bleaching with period $1/t_{12}$. Then, in the same way as a spatial grating
deflects a probe beam at an angle determined by the inverse ridge spacing, the spectral grating delays the response to
pulse~3, acting as a probe. The response delay equals $t_{12}$, the inverse period of the bleaching spectral modulation. 

In order to express an oscillating dipole in terms of the driving pulses, let us define the positive and negative
frequency components of the $i$-labeled driving field $E_i(\bm{r},t)$, centered at time $t_i$, as
\begin{align}\label{eq:envelope}
   E_i(\bm{r},t) & = \frac{1}{2}\left[\mathscr{A}_i(\bm{r},t)\exp(i\omega_Lt)+\text{c.c.}\right]
\\
                 & = \frac{1}{2}\left[\mathscr{E}_i(\bm{r},t-t_i)+\text{c.c.}\right]
\end{align}
where $\omega_L$ represents the pulse central frequency. Interaction with an optical dipole is characterized by the Rabi frequency:    
\begin{equation}
   \Omega_i(\bm{r},t)= \frac{\mu_{ab}\mathscr{E}_i(\bm{r},t)}{\hbar}
\end{equation} 
where $\mu_{ab}$ stands for the transition dipole moment. We also need the time-to-frequency Fourier transform of the
Rabi frequency, defined as
\begin{equation}
   \widetilde{\Omega}_i(\bm{r},\omega)=\int \Omega_i(\bm{r},t)\exp[-i\omega t]\ud t.
\end{equation}
This quantity is a dimensionless number and $\left|\widetilde{\Omega}_i^*(\bm{r},\omega_L)\right|$ represents the pulse
area.  Since, according to \eq{eq:envelope}, $\Omega_i(\bm{r},t)$ is centered at $t=0$,
$\widetilde{\Omega}_i(\bm{r},\omega)$ is a slowly varying function of $\omega$ over the pulse bandwidth.

In the weak field limit, when $\left|\widetilde{\Omega}_i^*(\bm{r},\omega_L)\right|\ll 1$, the dipole
$d(\bm{r},\omega_{ab},t)$, oscillating at position $\bm{r}$ at frequency $\omega_{ab}$, can be expressed to lowest order
in the three driving fields as~\cite{mossberg1982time}
\begin{multline}
   d(\bm{r},\omega_{ab},t)=i\frac{\mu_{ab}}{4}\exp\left[-\frac{t-t_3+t_{12}}{T_2}-\frac{t_{23}}{T_1}\right]
\\\times
   \left\{\widetilde{\Omega}_1^*(\bm{r},\omega_{ab})
         \widetilde{\Omega}_2(\bm{r},\omega_{ab})
         \widetilde{\Omega}_3(\bm{r},\omega_{ab})\exp\left[i\omega_{ab}\left(t-t_3-t_{12}\right)\right]
   \right.
\\\left.\vphantom{\widetilde{\Omega}_1^*(\bm{r},\omega_{ab})}
         -\text{c.c.}\right\}
\end{multline}
where $T_1$ represents the upper level lifetime. At time $t=t_3+t_{12}$ all the dipoles, irrespective of the transition
frequency value, are phased back together since $\omega_{ab}\left(t-t_3-t_{12}\right)$ vanishes, which results in the
photon echo emission.  Since the oscillating dipole at $\bm{r}$ is expressed in terms of the fields at the same
position, with no additional space dependence, this description applies not only to transparent, high optical quality
media, but also to scattering materials.

Similar expressions describe the various optical four-wave mixing processes. An important difference deserves to be
noticed, which is the absence of contribution proportional to
$\widetilde{\Omega}_1(\bm{r},\omega_{ab})\widetilde{\Omega}_2^*(\bm{r},\omega_{ab})\widetilde{\Omega}_3(\bm{r},\omega_{ab})$.
The extinction of this term is not related to any spatial phase matching condition but instead reflects
causality~\cite{mossberg1982time}.

\section{Physical model for photon echoes in disordered materials}\label{sec:model}

\subsection{Structure of the medium}

\begin{figure}[!htb]
   \centering
   \psfrag{E}[c]{$E_0$}
	\psfrag{e1}[c]{Scatterer}
   \psfrag{e2}[c]{\parbox{2cm}{Echo\\production}}
   \psfrag{e3}[c]{\parbox{2cm}{Scatterer and\\echo production}}
   \psfrag{L}[c]{$L$}
   \psfrag{x}[c]{$z$}
   \psfrag{y}[c]{$x$}
   \psfrag{a}[c]{(a)}
   \psfrag{b}[c]{(b)}
   \includegraphics[width=\linewidth]{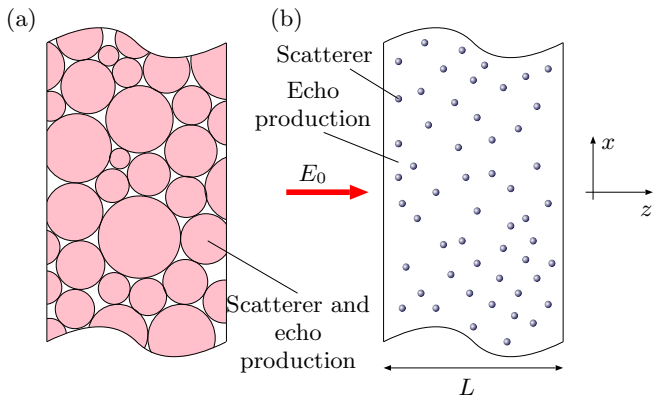}
   \caption{(a) Sketch of the experimentally investigated system, composed of assembled crystalline grains. (b) Model
   used in the theoretical and numerical studies. In the echo-generating continuous homogeneous medium, point scatterers
   are immersed at random, which gives rise to the multiple scattering effect. The particles are lying in a slab
   geometry of size $L$ which is convenient to derive simple analytical expressions.}
   \label{fig:slab}
\end{figure}

The polycrystalline powder in which photon echo has been observed~\cite{beaudoux2011} can be sketched as an ensemble of
contiguous, disorderly distributed, microscopic grains that contain the active echo-generating material
[\fig{fig:slab}\,(a)]. Successive reflection and refraction processes at the grain-walls result in the observed multiple
scattering effect.

Such a disordered succession of index-steps is difficult to model. Instead, we propose a much simpler scheme that
preserves the two leading features, namely echo generation and multiple scattering. We replace the original structure by
(1) an echo-generating continuous and homogeneous active medium, with (2) randomly embedded point-scatterers. This model
is illustrated in \fig{fig:slab}\,(b). One can switch from the disordered sample to the corresponding homogeneous slab
by just removing the scatterers. This offers an easy way to compare the signals in these two situations.

Moreover, to achieve large optical thickness numerically, we consider a two dimensional system and scalar waves (\ie the
electric field is oriented along the translational invariant direction $y$) embedded in a slab geometry with size $L$
along $z$ and infinite along $x$ as shown in \fig{fig:slab}. Translational invariance along $y$ results in substituting
the point scatterers with $N_s$ rectilinear, infinitely long, thin rods, randomly placed inside the system at positions
$\bm{r}_j$. With transverse size much smaller than the optical wavelength, the rods are assumed to behave as isotropic
scatterers. In addition, light is scattered elastically, without absorption in the rods.  This simplified model
does not permit a quantative comparison with the experiment of Ref.~\onlinecite{beaudoux2011} but contains all physical
ingredients (scattering and echo production) required to give physical insights into the existence of the strong
correlation between the driving fields and the echo beam.

\subsection{Coupled wave equations}\label{subsec:wave_equations}

Let $E^{(1,2,3)}(\bm{r},\omega)$ denote the three driving field spectral amplitudes, at position $\bm{r}$ and frequency
$\omega$. One also defines the exciting field $E_{\text{exc}}^{(1,2,3)}(\bm{r}_i,\omega)$ at rod position $\bm{r}_i$.
The latter corresponds to the field illuminating the scatterer, and is obtained by subtracting the scatterer emission
from the total field. The wave equation reads~\cite{lax1952}
\begin{multline}\label{eq:coupled_dipoles}
   E_{\text{exc}}^{(1,2,3)}(\bm{r}_i,\omega)=E_0(\bm{r}_i,\omega)
\\
      +k_0^2\alpha\sum_{\substack{j=1\\j\ne i}}^{N_s} G_0(\bm{r}_i-\bm{r}_j,\omega)E_{\text{exc}}^{(1,2,3)}(\bm{r}_j,\omega),
\end{multline}
where $k_0=\omega/c$ is the wave vector in vacuum, $\alpha$ represents the scatterer polarizability, and $G_0$ is the
free space Green function, connecting the field $E$ at any position inside the system to an electric-dipole point source $p$ lying at
position $\bm{r}_0$ by
\begin{equation}
   E(\bm{r},\omega)=\mu_0\omega^2 G_0(\bm{r}-\bm{r}_0,\omega)p.
\end{equation}
In a 2D scalar problem, $G_0$ is given by the isotropic function
\begin{equation}\label{eq:vacuum_green_function}
   G_0(\bm{r}-\bm{r}_0,\omega)=\frac{i}{4}\operatorname{H}_0^{(1)}(k_0|\bm{r}-\bm{r}_0|)
\end{equation}
where $\operatorname{H}_0^{(1)}$ denotes the Hankel function of first kind and zero order. 

In elastic, isotropic, scattering conditions, energy conservation leads to
\begin{equation}\label{eq:optical_theorem}
   k_0\im\alpha=\frac{k_0^3}{4}|\alpha|^2
\end{equation}
where the left-hand and right-hand sides, respectively, represent the extinction and scattering cross
sections~\cite{foldy1945,lax1951,lax1952}. Therefore, the cross section cannot exceed $4/k_0$, which corresponds to 
\begin{equation}\label{eq:alpha_max}
   \alpha=\alpha_{\text{max}}=\frac{4i}{k_0^2}
\end{equation}  

For the sake of simplicity, one derives the three driving fields from identical incident fields, denoted as $E_0$, and
given by plane waves at normal incidence
\begin{equation}\label{eq:plane_wave_def}
   E_0(\bm{r},\omega)=E_0\exp[ik_0z].
\end{equation}
Once the exciting fields are known, the electric driving fields can be calculated at any position inside or outside the medium thanks to the relation
\begin{multline}\label{eq:coupled_dipoles_field}
   E^{(1,2,3)}(\bm{r},\omega)=E_0(\bm{r},\omega)
\\
      +k_0^2\alpha\sum_{j=1}^{N_s} G_0(\bm{r}-\bm{r}_j,\omega)E_{\text{exc}}^{(1,2,3)}(\bm{r}_j,\omega).
\end{multline}

The echo is created by active atoms placed inside the host medium.
In the previous section, we have established that the source of the echo beam is given by
$\widetilde{\Omega}_1^*(\bm{r},\omega_{ab})\widetilde{\Omega}_2(\bm{r},\omega_{ab})\widetilde{\Omega}_3(\bm{r},\omega_{ab})$. Thus, the electric field of the echo signal can be cast in the form
\begin{multline}\label{eq:coupled_dipoles_echo}
   E_{\text{exc}}^{(4)}(\bm{r}_i,\omega)=
      k_0^2\chi \int G_0(\bm{r}_i-\bm{r}',\omega)E^{(1)*}(\bm{r}',\omega)
\\
      \qquad\qquad\qquad\times E^{(2)}(\bm{r}',\omega)E^{(3)}(\bm{r}',\omega)\ud\bm{r}'
\\
      +k_0^2\alpha\sum_{\substack{j=1\\j\ne i}}^{N_s}
      G_0(\bm{r}_i-\bm{r}_j,\omega)E_{\text{exc}}^{(4)}(\bm{r}_j,\omega).
\end{multline}
where $\chi$ is a constant describing the coupling between the driving fields and the echo beam.
Exactly as for the driving fields, the electric field of the echo at any position can be deduced from the relation
\begin{multline}\label{eq:coupled_dipoles_echo_field}
   E^{(4)}(\bm{r},\omega)=
      k_0^2\chi \int G_0(\bm{r}-\bm{r}',\omega)E^{(1)*}(\bm{r}',\omega)
\\
      \qquad\qquad\qquad\times E^{(2)}(\bm{r}',\omega)E^{(3)}(\bm{r}',\omega)\ud\bm{r}'
\\
      +k_0^2\alpha\sum_{j=1}^{N_s} G_0(\bm{r}-\bm{r}_j,\omega)E_{\text{exc}}^{(4)}(\bm{r}_j,\omega).
\end{multline}

This set of equations is overall reminiscent of previous descriptions of non-linearities in complex
systems~\cite{KRAVTSOV-1991,TOMITA-2004,SKIPETROV-2004,wellens2008}, and similarities with these works will be found all
along the following theory.

For a given spatial distribution of the $N_s$ scatterers, referred to as a \textit{configuration}, one has to solve the
$N_s$-linear-equation-set represented by \eq{eq:coupled_dipoles}. Then, with the help of \eq{eq:coupled_dipoles_field},
one can calculate the source term in \eq{eq:coupled_dipoles_echo}, which leads to the $N_s$ values of the exciting echo
field $E_{\text{exc}}^{(4)}(\bm{r}_j,\omega)$. Finally, substitution of $E_{\text{exc}}^{(4)}(\bm{r}_j,\omega)$ into
\eq{eq:coupled_dipoles_echo_field} determines the echo field anywhere, inside or outside the sample. The large size of
the system linear equations can be handled only through numerical computation.

\subsection{Configurational average}

The available experimental data are generally insufficient to define a specific configuration. Conversely, the
detailed field structure, as provided by the numerical solution, often exceeds the detector spatial, angular, or
temporal resolution. Therefore the experimentally accessible data, averaged over space and angle (in a rigid sample), or
time (in a fluid), are expected to coincide with statistical averages over all possible disordered configurations. Of
course, averaging washes out fine details, such as the speckle pattern of a fluctuating intensity emerging from a
disordered medium.

One can approach the statistical average numerically, by averaging the solutions over a set of different
configurations. More interestingly, in contrast with the single configuration problem, statistical average is accessible
analytically. As will become clear in the following, the analytical solution not only saves computation time, but also
brings physical insight into the observable quantities.

In the next two sections we shall adapt the available tools to echo generation in disordered media. The numerical
solution, discussed in Sec.~\ref{sec:numerics}, will serve to validate the analytical procedure.   

\section{The multiple scattering theory for the driving fields}\label{sec:driving_theory}

In this section, we derive the average amplitude and intensity of the driving fields, as well as another quantity called
the ladder operator, in order to get the necessary building blocks to obtain the echo signal. Since this is a textbook
formalism, we only summarize the key steps. The interested reader may refer to
Refs.~\onlinecite{ROSSUM-1999,MONTAMBAUX-2007} to find more details.

\subsection{Average field}

Let us first compute the average field. For that purpose, we combine \eqs{eq:coupled_dipoles} and
\eqref{eq:coupled_dipoles_field} to obtain a cluster expansion of the driving fields. Omitting the exponents $(1,2,3)$
and the frequency $\omega$ related to the driving fields for the sake of simplicity, we get~\cite{FRISCH-1967}
\begin{multline}\label{eq:cluster_field}
   E(\bm{r})=E_0(\bm{r})
      +k_0^2\alpha\sum_{i=1}^{N_s} G_0(\bm{r}-\bm{r}_i)E_0(\bm{r}_i)
\\
      +k_0^2\alpha\sum_{i=1}^{N_s} G_0(\bm{r}-\bm{r}_i)k_0^2\alpha\sum_{\substack{j=1\\j\ne i}}^{N_s} G_0(\bm{r}_i-\bm{r}_j)E_0(\bm{r}_j)
      +\ldots
\end{multline}
Averaging \eq{eq:cluster_field} over the configurations of the disorder leads to a closed and exact equation called the
Dyson equation~\cite{DYSON-1949,DYSON-1949-1}. In the following, the discussion is restricted to the Independent
Scattering Approximation (ISA), where all the scattering events along a scattering sequence are statistically
independent. The ISA is valid in a dilute medium, and the corresponding condition will be elucidated soon. In this
limit, Dyson equation reads as
\begin{equation}\label{eq:dyson_field}
   \bra E(\bm{r})\ket=E_0(\bm{r})
      +\rho_s k_0^2\alpha\int G_0(\bm{r}-\bm{r}')\bra E(\bm{r}')\ket\ud\bm{r}'
\end{equation}
where the brackets $\bra\ldots\ket$ denote the statistical average and $\rho_s$ is the density of scatterers. 

Formal iterative solution of \eq{eq:dyson_field} leads to
\begin{equation}\label{eq:dyson_formal_sol}
   \bra E(\bm{r})\ket=E_0(\bm{r})
      +\rho_s k_0^2\alpha\int \bra G(\bm{r}-\bm{r}')\ket E_0(\bm{r}')\ud\bm{r}'
\end{equation}
where
\begin{multline}\label{eq:dyson_green}
   \bra G(\bm{r}-\bm{r}_0)\ket=G_0(\bm{r}-\bm{r}_0)
\\
      +\rho_s k_0^2\alpha\int G_0(\bm{r}-\bm{r}')\bra G(\bm{r}'-\bm{r}_0)\ket\ud\bm{r}'.
\end{multline}

Hence, \eq{eq:dyson_formal_sol} expresses $\bra E(\bm{r})\ket$ in terms of $E_0$ and of the average Green function $\bra
G\ket$, which can be obtained by solving \eq{eq:dyson_green}. Actually, \eq{eq:dyson_green} is the Dyson equation for
the average Green function $\bra G(\bm{r}-\bm{r}_0)\ket$, the average field radiated at position $\bm{r}$ by a point
source, located at $\bm{r}_0$.

To solve \eq{eq:dyson_green} we assume a bulk geometry, ignoring the finite size of the actual slab. In this framework,
Fourier transforming \eq{eq:dyson_green} leads to:
\begin{equation}\label{eq:dyson_green_fourier}
   \bra G(\bm{k})\ket=G_0(\bm{k})+G_0(\bm{k})\rho_s k_0^2\alpha\bra G(\bm{k})\ket.
\end{equation}
Substituting the Fourier transform of the free-space Green function
\begin{equation}\label{eq:vacuum_green_function_fourier}
   G_0(\bm{k})=\left(k^2-k_0^2\right)^{-1} 
\end{equation}
into \eq{eq:dyson_green_fourier}, we readily get
\begin{equation}\label{eq:dyson_green_fourier_sol}
   \bra G(\bm{k})\ket=\left(k^2-\keff^2\right)^{-1}.
\end{equation}
where
\begin{equation}\label{eq:keff}
   \keff=k_0\sqrt{1+\rho_s\alpha}.
\end{equation}

Except for the substitution of $k_0$ with $\keff$, $G_0(\bm{k})$ and $\bra G(\bm{k})\ket$ are expressed in the same way.
As a consequence, inverse Fourier transform of \eq{eq:dyson_green_fourier_sol} leads to
\begin{equation}\label{eq:average_green_function}
   \bra G(\bm{r}-\bm{r}_0)\ket
      =\frac{i}{4}\operatorname{H}_0^{(1)}(\keff|\bm{r}-\bm{r}_0|),
\end{equation}
which can be compared to the vacuum counterpart given by \eq{eq:vacuum_green_function}. The average field propagates in
an effective system with an effective permittivity, the imaginary part of which describes the attenuation due to
scattering (loss by scattering). Indeed
\begin{equation}\label{eq:keff_expansion} 
   \keff \sim k_0+\frac{i}{2\ell}
\end{equation}
where the scattering mean-free path $\ell$ (average distance between two consecutive scattering events) is given by
\begin{equation}\label{eq:ell}
  1/\ell=\rho_s k_0\im\alpha.
\end{equation}

At this stage we are able to explicit the ISA condition in a dilute system as $k_0\ell\gg1$. 

In the slab geometry, under illumination by a plane wave $E_0(z)$ at normal incidence to the interfaces,
\eq{eq:dyson_formal_sol} reduces to
\begin{equation}\label{eq:dyson_formal_sol_1D}
   \bra E(z)\ket=E_0(z)
      +(\keff^2-k_0^2)\int_0^L \bra G(z-z')\ket E_0(z')\ud z'
\end{equation}
where
\begin{align}\label{eq:Green_1D}
   \bra G(z)\ket & = \frac{i}{4}\int_{-\infty}^\infty\operatorname{H}_0^{(1)}(\keff\sqrt{x^2+z^2})\ud x
\\
                 & = \frac{i}{2\keff}\exp\left[i\keff|z|\right]
\end{align}
is the 1D average Green function.

Substituting $E_0(z)$ given by \eq{eq:plane_wave_def} into \eq{eq:dyson_formal_sol_1D}, one readily obtains: 
 \begin{equation}\label{eq:average_field}
   \bra E(z)\ket=E_0\exp\left[i\keff z\right]
\end{equation}
in the ISA conditions.
 
The corresponding intensity (often called ballistic or coherent intensity) is given by
\begin{equation}\label{eq:ballistic_intensity}
   I_B(z)=\left|\bra E(z)\ket\right|^2 = I_0\exp\left[-\frac{z}{\ell}\right].
\end{equation}
From this, one can define the optical thickness as the ratio $b=L/\ell$, in terms of which the relative power,
ballistically transmitted by the system, can be expressed as $T_B=\exp(-b)$.

It is usual in the multiple scattering theory to have a simple representation of iterative equations in terms of
diagrams. For \eq{eq:dyson_green}, it reads
\begin{multline}
   \bra G(\bm{r}-\bm{r}_0)\ket=
      \begin{diagc}{6}
         \ginc{0}{6}
      \end{diagc}+
      \begin{diagc}{12}
         \ginc{0}{6}
         \ginc{6}{12}
         \particule{6}
      \end{diagc}
\\
      +
      \begin{diagc}{18}
         \ginc{0}{6}
         \ginc{6}{12}
         \ginc{12}{18}
         \particule{6}
         \particule{12}
      \end{diagc}+
      \ldots   
\end{multline}
where circles and solid lines denote scattering events and free-space Green functions $G_0$ respectively.

\subsection{Average intensity}\label{subsec:average_intensity}

The same work can be carried out to compute the average intensity. The field correlation $\bra
E(\bm{r})E^*(\bm{\uprho})\ket$, coinciding with the average intensity when $\bm{r}=\bm{\uprho}$, is driven by the
Bethe-Salpeter equation~\cite{RYTOV-1989,APRESYAN-1996}, which, in the dilute system approximation, reduces to
\begin{multline}\label{eq:bethe_salpeter_approx}
   \bra I(\bm{r})\ket=\left|\bra E(\bm{r})\ket\right|^2
\\
      +\frac{4k_0}{\ell}\int \left|\bra G(\bm{r}-\bm{r}')\ket\right|^2\bra I(\bm{r}')\ket\ud\bm{r}'.
\end{multline}

The iterative solution to that equation can be expanded as a series of diagrams: 
\begin{equation}
   \bra I(\bm{r})\ket = 
   \begin{ddiag}{4}
      \eemoy{0}{4}{-3}
      \eemoy{0}{4}{3}
   \end{ddiag}+
   \begin{ddiag}{10}
      \ggmoy{0}{5}{-3}
      \ggmoy{0}{5}{3}
      \eemoy{5}{10}{-3}
      \eemoy{5}{10}{3}
      \iidentique{5}{3}{5}{-3}
      \pparticule{5}{-3}
      \pparticule{5}{3}
   \end{ddiag}+
   \begin{ddiag}{16}
      \ggmoy{0}{5}{-3}
      \ggmoy{0}{5}{3}
      \ggmoy{5}{11}{-3}
      \ggmoy{5}{11}{3}
      \eemoy{11}{16}{-3}
      \eemoy{11}{16}{3}
      \iidentique{5}{3}{5}{-3}
      \iidentique{11}{3}{11}{-3}
      \pparticule{5}{-3}
      \pparticule{5}{3}
      \pparticule{11}{-3}
      \pparticule{11}{3}
   \end{ddiag}+\ldots
\end{equation}
where the upper (lower) line corresponds to the field (its conjugate) respectively. Thick solid lines correspond to the average Green functions and thick dashed lines denote average fields. The circles represent the scattering events, which are joined by vertical lines since they occur at the same position, in the same order, for both fields. The resulting characteristic shape is known as a ladder diagram.

Not only does that diagram expansion represent a convenient mathematical tool, but it also conveys a physical
picture for the averaged intensity propagation through a disordered medium. Indeed, as illustrated by this diagram,
statistical average washes out most of the contributions to $\bra I\ket $ at position $\bm{r}$ -- those affected by
the erratic spatial phase factors that build up when $E$ and $E^*$ follow different paths, and strongly depend on the
path details. Only survive the scattering sequences where both fields $E$ and $E^*$ follow the same path, with the same
scatterers located at the same positions. As it will soon become clear, that drastic selection results in speckle
structure erasure. Averaging over a spatial region, with volume larger than $\lambda^3$, for a given and fixed scatterer
distribution, is expected to achieve the same scattering path selection as statistical averaging over scatterer
distributions.

With the help of the ballistic intensity defined in \eq{eq:ballistic_intensity}, one readily casts
\eq{eq:bethe_salpeter_approx} in the form
\begin{multline}\label{eq:bethe_salpeter_ladder_ell}
   I_D(\bm{r})-\frac{4k_0}{\ell}\int \left|\bra G(\bm{r}-\bm{r}')\ket\right|^2 
      I_D(\bm{r}')\ud\bm{r}' 
	\\ 
			= \frac{4k_0}{\ell}\int \left|\bra G(\bm{r}-\bm{r}')\ket\right|^2 
      I_B(\bm{r}')\ud\bm{r}'.
\end{multline}
where $I_D(\bm{r})=\bra I(\bm{r})\ket-I_B(\bm{r})$ represents the diffuse intensity. Deep inside the medium, at
distances $\gg\ell$ from the interfaces, $\left|\bra G(\bm{r}-\bm{r}')\ket\right|^2$ is given by
\eq{eq:average_green_function}. In this region, one may simplify the left-hand side of
\eq{eq:bethe_salpeter_ladder_ell}, observing that the spatial frequency spectrum of $I_D(\bm{r})$ is much narrower than
that of $\left|\bra G(\bm{r}-\bm{r}')\ket\right|^2$. Hence one may replace the latter function Fourier transform by its
second order Taylor expansion, which leads to
\begin{multline}\label{eq:bethe_salpeter_left_hd_side}  
   I_D(\bm{r})-\frac{4k_0}{\ell}\int \left|\bra G(\bm{r}-\bm{r}')\ket\right|^2 
      I_D(\bm{r}')\ud\bm{r}'
\\			
   =-\frac{\ell^2}{2}\Delta I_D(\bm{r})
\end{multline}
The right-hand side in \eq{eq:bethe_salpeter_ladder_ell}, operating as a source term, vanishes far from the interfaces,
in the region where \eq{eq:average_green_function} is valid. Hence, according to \eq{eq:bethe_salpeter_left_hd_side},
\eq{eq:bethe_salpeter_ladder_ell} reduces to $\Delta I_D(\bm{r})=0$, which conveys no information on $I_D(\bm{r})$
build-up from ballistic intensity. Closer to the input interface, the source term no longer vanishes but the bulk
approximation, ignoring the finite size of the slab, no longer applies. However, numerical simulations appear to be
consistent with \eq{eq:bethe_salpeter_left_hd_side}, provided the right-hand side of this equation is replaced with
$I_B(\bm{r})$. 
 
The resulting diffusion equation~\cite{MONTAMBAUX-2007}, now considered to be valid throughout the medium, reads as
\begin{equation}\label{eq:diffusion}
   -\frac{\ell^2}{2}\Delta I_D(\bm{r})=I_B(\bm{r}).
\end{equation}

That equation is complemented by two boundary conditions, assessing the absence of incoming diffuse intensity through both interfaces:
\begin{align}\label{eq:boundary_cond}
   I_D(z=0)-z_0\left.\frac{\partial I_D(z)}{\partial z}\right|_{z=0} & =0,
\\
   I_D(z=L)+z_0\left.\frac{\partial I_D(z)}{\partial z}\right|_{z=L} & =0,
\end{align}
where $z_0$, the so called extrapolation length~\cite{ISHIMARU-1997}, is on the order of $\ell$. Let $F(z)$ represent any solution of equation
\begin{equation}\label{eq:diffusion_with_z}
   \frac{\ell^2}{2}F''(z)=I_0\exp\left[-\frac{z}{\ell}\right].
\end{equation}
Then, the solution of the diffusion equation, consistent with the boundary conditions, reads
\begin{multline}\label{eq:diffusion_eq_solution}
   I_D(z)=\frac{(z+z_0)F(L)+(L+z_0-z)F(0)}{L+2z_0}
\\
   +\frac{z_0(z+z_0)F'(L)-z_0(L+z_0-z)F'(0)}{L+2z_0}-F(z).
\end{multline}
Finally the solution of \eq{eq:diffusion} reads
\begin{equation}\label{eq:diffuse_intensity}
   I_D(z)
      =2I_0\left[\left(1+\frac{z_0}{\ell}\right)\frac{L+z_0-z}{L+2z_0}-\exp\left(-\frac{z}{\ell}\right)\right].
\end{equation}
The factor $z_0/\ell$ in this equation makes $I_D(z)$ sensitive to $z_0$ at any depth in the medium.

\begin{figure}[!htb]
   \centering
   \psfrag{z}[c]{$z/\ell$}
   \psfrag{AAAAAA}[bl]{$I_D/I_0$ numerical}
   \psfrag{BBBBBB}[bl]{$I_D/I_0$ analytical}
   \psfrag{CCCCCC}[bl]{$I_B/I_0$}
   \includegraphics[width=\linewidth]{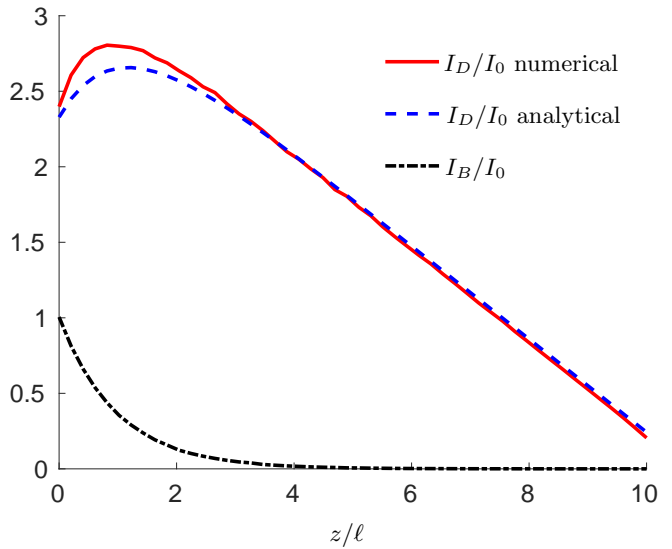}
   \caption{(Color online) Ballistic ($I_B$, black dash-dotted line) and diffuse ($I_D$, blue dashed line and red solid
   line) intensity as a function of the depth $z$ inside the slab for $b=10$ and $k\ell=40$.}
   \label{fig:ballistic_diffuse}
\end{figure}

The variations of $I_B$ and $I_D$ with $z$ are plotted in \fig{fig:ballistic_diffuse}. We have taken $z_0=\pi\ell/4$, a
standard value for a 2D problem~\cite{MONTAMBAUX-2007}. We may notice the very fast decay of $I_B$ on a typical
length given by $\ell$, and the slower decay of $I_D$. This plot represents a stationary state, where the sample,
continuously fed by the incident plane wave, re-emits all that energy through the interfaces. Although
\eq{eq:diffuse_intensity} is valid only within the slab boundaries, the spatial intensity distribution, as represented
in \fig{fig:ballistic_diffuse}, suggests that most of the incoming flux is scattered in backward direction through the
input interface.   

According to \eqs{eq:ballistic_intensity} and \eqref{eq:diffuse_intensity}, the average intensity
reads
\begin{equation}\label{eq:average_intensity}
   \bra I(z)\ket
      =I_0\left[2\left(1+\frac{z_0}{\ell}\right)\frac{L+z_0-z}{L+2z_0}-\exp\left(-\frac{z}{\ell}\right)\right].
\end{equation}

\subsection{Ladder operator}

In the same way as we have defined the average Green function for the average field, we may define a Green function for
the average intensity. Called the ladder operator, this quantity can be represented by the diagram
\begin{equation}
   L(\bm{r},\bm{r}_0) = 
   \begin{ddiag}{2}
      \iidentique{1}{3}{1}{-3}
      \pparticule{1}{-3}
      \pparticule{1}{3}
   \end{ddiag}+
   \begin{ddiag}{8}
      \ggmoy{1}{7}{-3}
      \ggmoy{1}{7}{3}
      \iidentique{1}{3}{1}{-3}
      \pparticule{1}{-3}
      \pparticule{1}{3}
      \iidentique{7}{3}{7}{-3}
      \pparticule{7}{-3}
      \pparticule{7}{3}
   \end{ddiag}+
   \begin{ddiag}{14}
      \ggmoy{1}{7}{-3}
      \ggmoy{1}{7}{3}
      \ggmoy{7}{13}{-3}
      \ggmoy{7}{13}{3}
      \iidentique{1}{3}{1}{-3}
      \pparticule{1}{-3}
      \pparticule{1}{3}
      \iidentique{7}{3}{7}{-3}
      \pparticule{7}{-3}
      \pparticule{7}{3}
      \iidentique{13}{3}{13}{-3}
      \pparticule{13}{-3}
      \pparticule{13}{3}
   \end{ddiag}+\ldots
\end{equation}
which analytically gives
\begin{equation}\label{eq:bethe_salpeter_ladder}
   L(\bm{r},\bm{r}_0)=\frac{4k_0}{\ell}\delta(\bm{r}-\bm{r}_0)+\frac{4k_0}{\ell}
      \int \left|\bra G(\bm{r}-\bm{r}')\ket\right|^2L(\bm{r}',\bm{r}_0)\ud\bm{r}'.
\end{equation}

In large systems ($L\gg\ell$), \eq{eq:bethe_salpeter_ladder} reduces to
\begin{equation}\label{eq:diffusion_ladder}
   -\frac{\ell^2}{2}\Delta L(\bm{r},\bm{r}_0)=\frac{4k_0}{\ell}\delta(\bm{r}-\bm{r}_0).
\end{equation}
Expressed in terms of $\left|\bra G(\bm{r}-\bm{r}')\ket\right|^2$, just as \eq{eq:bethe_salpeter_approx}, the ladder
operator does not help to handle source terms such as $I_B(\bm{r})$, which are strongly confined to the close vicinity
of the interfaces. However, as will be seen soon, it proves helpful to deal with slowly varying sources terms, spreading
all over the medium. 

\section{The multiple scattering theory for the echo signal}\label{sec:echo_theory}

This section is the original part of the study. We intend to analytically derive the statistical average of the echo
field and intensity and to express the echo field correlation with one of the driving fields. According to
Ref.~\onlinecite{beaudoux2011}, the amplitude of this correlation can be large, even in the multiple scattering regime. The
following derivation will help to understand the origin of this strong correlation.

\subsection{Average echo field}

Along the lines of the above summarized multiple scattering theory, we have to identify the most important diagrams in
the context of photon echo physics. Let us focus first on the average echo field. According to
\eqs{eq:coupled_dipoles_echo} and \eqref{eq:coupled_dipoles_echo_field}, the echo signal is created at any position in
the host medium and is given by $E^{(1)*}(\bm{r}',\omega)E^{(2)}(\bm{r}',\omega)E^{(3)}(\bm{r}',\omega)$. To construct
the average echo field, we let the  average intensity and the average field merge at $\bm r'$. The resulting signal
propagation from $\bm r'$ to $\bm r$ is carried out by the average Green function. That scheme is represented by the
following diagram structure:
\begin{equation}
   \bra E^{(4)}(\bm{r})\ket = 
   \begin{dddiag}{24}
      \ggmoy{0}{6}{0}
      \eemoy{6}{24}{0}
      \gggmoy{6}{12}{0}{12}
      \gggmoy{6}{12}{0}{6}
      \ggmoy{12}{18}{12}
      \ggmoy{12}{18}{6}
      \eemoy{18}{24}{12}
      \eemoy{18}{24}{6}
      \iidentique{12}{6}{12}{12}
      \iidentique{18}{6}{18}{12}
      \pparticule{12}{12}
      \pparticule{18}{12}
      \pparticule{12}{6}
      \pparticule{18}{6}
      \nnonlineaire{6}{0}
   \end{dddiag}+\ldots
\end{equation}
where the square represents echo generation; the middle line corresponds to $E^{(1)*}$; the upper and lower lines may
respectively refer to $E^{(2)}$ and $E^{(3)}$, or to $E^{(3)}$ and $E^{(2)}$.  All the significant contributions to
$\bra E^{(4)}(\bm{r})\ket$ share the same structure, with different numbers of scattering events on each branch.  Due to
the possible permutation of $E^{(2)}$ and $E^{(3)}$, each diagram should be counted twice. 

Finally, in quite the same way as the incident average field [see \eq{eq:dyson_formal_sol_1D}], $\bra
E^{(4)}(\bm{r})\ket$ can be expressed analytically as follows:  
\begin{equation}\label{eq:average_echo_field_eq}
   \bra E^{(4)}(z)\ket=2k_0^2\chi\int_{0}^L \bra G(z-z')\ket \bra I(z')\ket
      \bra E(z')\ket\ud z',
\end{equation}
still in a dilute system with $k_0\ell\gg 1$. In the large optical thickness limit, with $b\gg 1$, the integral upper
bound can be changed into $z$, without significant deviations except at the very beginning of the slab ($z<\ell$).
Using \eqs{eq:average_field} and \eqref{eq:average_intensity}, we finally get
\begin{multline}\label{eq:average_echo_field}
   \bra E^{(4)}(z)\ket=ik_0\ell\chi E_0I_0\exp\left[i\keff z\right]
\\\times
      \left\{\frac{2}{\ell}\left(1+\frac{z_0}{\ell}\right)\frac{(L+z_0)z-z^2/2}{L+2z_0}+\exp\left[-\frac{z}{\ell}\right]-1\right\}.
\end{multline}
In terms of intensity, this gives
\begin{equation}\label{eq:ballistic_echo_intensity}
   I_B^{(4)}(z)=\left|\bra E^{(4)}(z)\ket\right|^2.
\end{equation}

These quantities can be compared with their equivalents in a homogeneous slab with the same active atom
concentration. To deprive the slab from all the scattering centers, we just replace the average Green function,
intensity and field in \eq{eq:average_echo_field_eq} by their counterparts for an homogeneous medium, which gives:
\begin{equation}\label{eq:echo_field_hom}
   E_{\text{hom}}^{(4)}(z) =k_0^2\chi\int_0^L G_0(z-z')I_0(z')E_0(z')\ud z'.
\end{equation}
Provided $k_0z\gg 1$, the corresponding homogeneous echo field and intensity reduce to:
\begin{align}\label{eq:echo_field_hom_app}
   E_{\text{hom}}^{(4)}(z) & = \frac{ik_0\chi E_0 I_0 z}{2}\exp[ik_0z],
\\\label{eq:echo_intensity_hom}
   \text{and } I_{\text{hom}}^{(4)}(z) & = \frac{k_0^2\chi^2 I_0^3z^2}{4}.
\end{align}
It should be pointed out that the bulk geometry approximation we have been using, imposing the large optical depth
condition $L\gg\ell$, forbids any continuous transition from a disordered to a homogeneous medium, for example
by continuously increasing the scattering mean free path $\ell$.    

\begin{figure}[!htb]
   \centering
   \psfrag{z}[c][t]{$z/\ell$}
   \psfrag{AAAAAA}[l][l]{$I_B^{(4)}/I^{(4)}_{\text{hom}}(L)$ numerical}
   \psfrag{BBBBBB}[l][l]{$I_B^{(4)}/I^{(4)}_{\text{hom}}(L)$ analytical}
   \psfrag{CCCCCC}[l][l]{$I^{(4)}_{\text{hom}}/I^{(4)}_{\text{hom}}(L)$}
   \includegraphics[width=\linewidth]{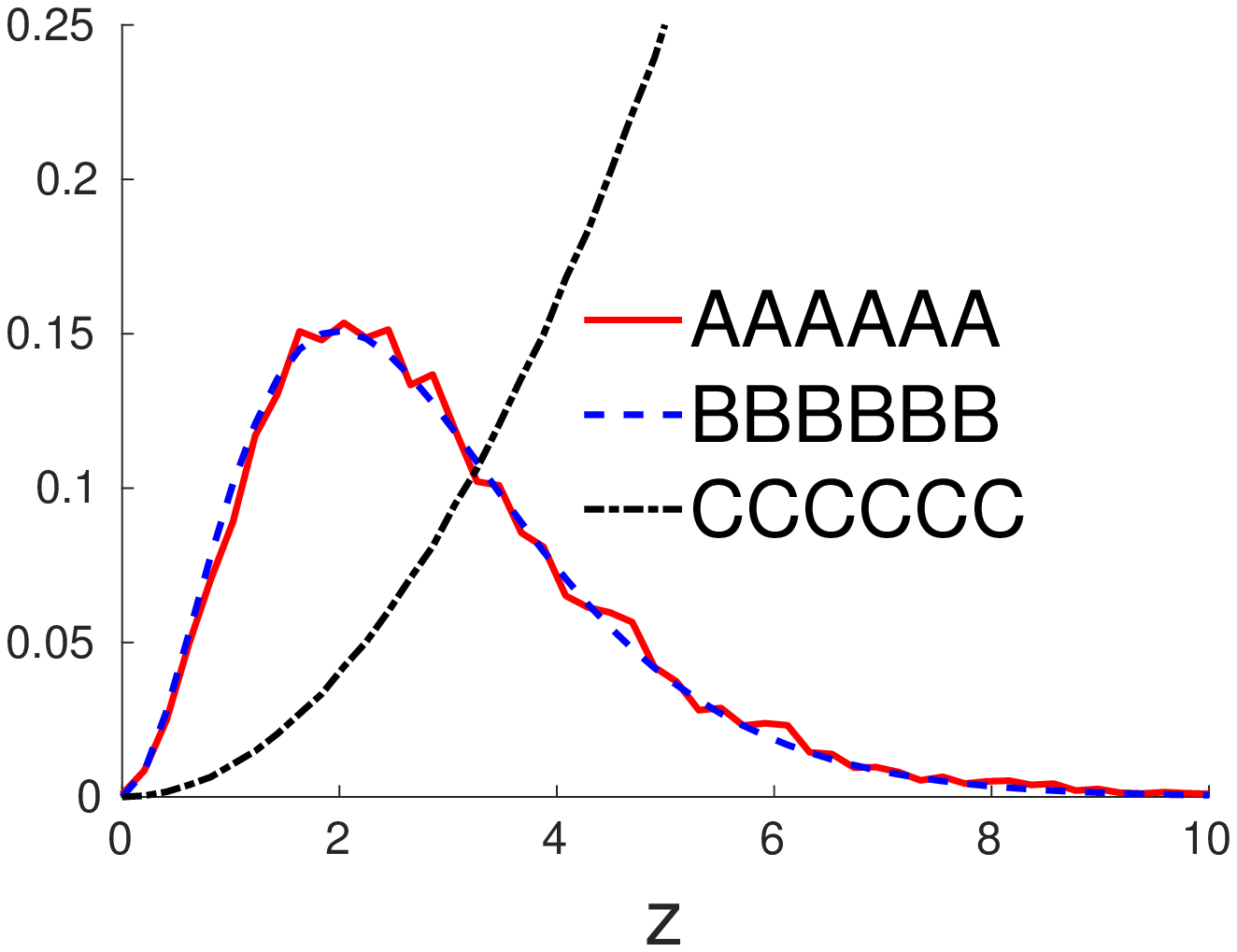}
   \caption{(Color online) Ballistic intensity of the echo $I_B^{(4)}=\left|\bra E^{(4)}\ket\right|^2$ as a function of the
   depth $z/\ell$ inside the slab (numerical computation: red solid line; analytical solution: blue dashed line); echo
   intensity in a homogeneous slab $I^{(4)}_{\text{hom}}$ (black dash-dotted line); $b=10$ and $k_0\ell=40$.}
   \label{fig:ballistic_echo_intensity}
\end{figure}

In \fig{fig:ballistic_echo_intensity}, we have plotted $I_B^{(4)}(z)$ and $I_{\text{hom}}^{(4)}(z)$, both normalized to
$I_{\text{hom}}^{(4)}(L)$, the echo intensity at the exit interface of a homogeneous slab. Close to the input,
$I_B^{(4)}(z)$ and $I_{\text{hom}}^{(4)}(z)$ exhibit the same parabolic variation with $z$, which is the signature of
spatially coherent buildup of the echo grows. In that region, the much faster growth of echo intensity in the disordered medium
reflects the incident energy confinement near the input side of the slab. However, while the signal intensity grows
quadratically in the homogeneous slab, a maximum is reached at $z\simeq 2\ell$ in the disordered medium,
followed by a fast decrease with $z$. Scattering affects both the echo generation and propagation. On the one hand, the
driving field $\bra E(z)\ket$ drops with $z$, reducing contributions to the signal deeper into the slab. On the other
hand, the ballistic component of the echo is attenuated as it propagates through the medium, feeding its diffuse part.

\subsection{Correlation of photon echoes with the driving fields}

We shall now present a key stage in our investigation, namely the calculation of the correlation function
\begin{equation}\label{eq:echo_correlation_def}
   C^{(4)}(\bm{r})=\bra E(\bm{r})E^{(4)*}(\bm{r})\ket,
\end{equation}
meaning the correlation of one driving field with the photon echo signal. This quantity is central in this work, first because
it has been accessed to experimentally~\cite{beaudoux2011}, second because its observed large amplitude represents a
counterintuitive result. Indeed the driving fields and the non-linear signal are expected to develop very different
distorted wavefronts as they travel through the disordered medium, which should hamper any correlation buildup.

One easily disposes of the ballistic component $C_B^{(4)}(z)=\bra E(z)\ket\bra E^{(4)*}(z)\ket$, with the help of
\eqs{eq:average_field} and \eqref{eq:average_echo_field}. This contribution dies out at short distance from the input
interface. Expressing the diffuse component $C_D^{(4)}(z)=C^{(4)}(z)-C_B^{(4)}(z)$ is more challenging. Proceeding along
the lines of the calculation of $I_D(z)$, we only retain the diagrams where the two participant fields follow the same
sequence, undergoing scattering events in the same order at the same positions. More precisely, the propagation path is
first followed by two incoming fields, one acting as a reference, the other as a driving field. They travel together up
to an interaction point where the driving fields disappear, giving birth to an echo. From that point on, the echo and
the reference field progress side by side along the same path. At each interaction point, the material response radiates
in all directions, but all these contributions are expected to be accounted for by summation over the different paths.
In the resulting diagram
\begin{multline}\label{eq:echo_correlation_diag}
   C_D^{(4)}(\bm{r})=
\\
   \begin{ddddiag}{36}
      \rput(6,6){$\bm{r}'$}
      \rput(12,6){$\bm{r}''$}
      \rput(18,0){$\bm{\uprho}$}
      \rput(24,6){$\bm{r}'''$}
      \ggmoy{0}{6}{-3}
      \ggmoy{6}{12}{-3}
      \ggmoy{12}{18}{-3}
      \ggmoy{18}{24}{-3}
      \ggmoy{24}{30}{-3}
      \eemoy{30}{36}{-3}
      \ggmoy{0}{6}{3}
      \ggmoy{6}{12}{3}
      \ggmoy{12}{24}{3}
      \ggmoy{24}{30}{3}
      \eemoy{30}{36}{3}
      \iidentique{6}{-3}{6}{3}
      \iidentique{12}{-3}{12}{3}
      \iidentique{24}{-3}{24}{3}
      \iidentique{30}{-3}{30}{3}
      \pparticule{6}{-3}
      \pparticule{12}{-3}
      \pparticule{24}{-3}
      \pparticule{30}{-3}
      \pparticule{6}{3}
      \pparticule{12}{3}
      \pparticule{24}{3}
      \pparticule{30}{3}
      \gggmoy{18}{24}{-3}{-9}
      \gggmoy{18}{24}{-3}{-15}
      \ggmoy{24}{30}{-9}
      \eemoy{30}{36}{-9}
      \ggmoy{24}{30}{-15}
      \eemoy{30}{36}{-15}
      \iidentique{24}{-15}{24}{-9}
      \iidentique{30}{-15}{30}{-9}
      \pparticule{24}{-9}
      \pparticule{30}{-9}
      \pparticule{24}{-15}
      \pparticule{30}{-15}
      \nnonlineaire{18}{-3}
   \end{ddddiag}+\ldots
\end{multline}
both propagations up to $\bm{r'''}$ and $\bm{\uprho}$ are described by the average intensity $\bra I\ket$, while the
side-by-side progression of the echo field from $\bm{r''}$ to $\bm{r'}$ is conveyed by the ladder
operator $L$. The box from $\bm{r'''}$ to $\bm{r''}$, which contains the conversion of the driving fields into the echo,
is given by
\begin{multline}
   K(\bm{r}'',\bm{r}''')=k_0^2\chi\int\bra G(\bm{r}''-\bm{r}''')\ket\bra G^*(\bm{r}''-\bm{\uprho})\ket
\\\times
   \bra I(\bm{\uprho})\ket\bra G^*(\bm{\uprho}-\bm{r}''')\ket\ud\bm{\uprho}.
\end{multline}
Putting all pieces together, one gets 
\begin{multline}\label{eq:diffuse_echo_correlation}
   C_D^{(4)}(\bm{r})=\frac{8k_0}{\ell}\int\left|\bra G(\bm{r}-\bm{r}')\ket\right|^2 L(\bm{r}',\bm{r}'')
      K(\bm{r}'',\bm{r}''')
\\\times
      \bra I(\bm{r}''')\ket \ud\bm{r}'\ud\bm{r}''\ud\bm{r}'''
\end{multline}
where a factor of two takes into account the driving field permutation. 

The origin of the correlation strength is all contained in the diagram~\eqref{eq:echo_correlation_diag}, and can be
realized already, without further calculation. Actually, the correlation buildup appears to be exactly as selective as
the diffuse intensity propagation scheme, discussed in Sec.~\ref{subsec:average_intensity}. In both cases one may
neglect all the contributions containing a spatial phase shift, only keeping the single-path diagrams. Hence both the
diffuse correlation and the diffuse intensity survive in the same way, traveling along the same
paths through the disordered medium.

According to \eq{eq:average_intensity}, the average intensity varies slowly in a large system (\ie $b\gg 1$), deep
inside the medium (\ie $z\gg \ell$). The same statement can be formulated for the ladder.  As the average Green function
scales typically with the scattering mean-free path $\ell$, $\bra I(\bm{\uprho})\ket$ and $L(\bm{r}',\bm{r}'')$ can be
replaced by $\bra I(\bm{r}''')\ket$ and $L(\bm{r},\bm{r}''')$ respectively in the above integrals. Moreover, using the
well-known identity~\cite{MONTAMBAUX-2007}
\begin{equation}\label{eq:identity}
   \frac{\ell}{k_0}\im\bra G(\bm{r}-\bm{r}'')\ket
      =\int \bra G(\bm{r}-\bm{r}')\ket\bra G^*(\bm{r}'-\bm{r}'')\ket\ud\bm{r}'
\end{equation}
for $\bm{r}''=\bm{r}$, we finally obtain
\begin{equation}\label{eq:diffuse_echo_correlation_ladder}
   C_D^{(4)}(\bm{r})=2\mathcal{K}
      \int L(\bm{r},\bm{r}''')\bra I(\bm{r}''')\ket^2 \ud\bm{r}'''
\end{equation}
with
\begin{multline}
   \mathcal{K}=k_0^2\chi\int\bra G(\bm{r}''-\bm{r}''')\ket\bra G^*(\bm{r}''-\bm{\uprho})\ket
\\
   \times\bra G^*(\bm{\uprho}-\bm{r}''')\ket\ud\bm{\uprho}\ud\bm{r}''.
\end{multline}  
Making use again of \eq{eq:identity}, we can simplify $\mathcal{K}$ into
\begin{equation}\label{eq:factor_k}
   \mathcal{K}
      =k_0\ell\chi\int \im\left[\bra G(\bm{\uprho})\ket\right]\bra G^*(\bm{\uprho})\ket\ud\bm{\uprho},
\end{equation}
which, for a dilute medium, reduces to
\begin{align}\label{eq:factor_k_final}
   \mathcal{K}
      & =\frac{k_0\ell\chi}{2i}\int\left|\bra G(\bm{\uprho})\ket\right|^2\ud\bm{\uprho}
\\
      & =\frac{\ell^2\chi}{2i}\im\left[\bra G(0)\ket\right]=\frac{\ell^2\chi}{8i}.
\end{align}

Applying the Laplace operator to \eq{eq:diffuse_echo_correlation_ladder}, and making use of \eq{eq:diffusion_ladder}, one finally obtains
\begin{equation}\label{eq:diffuse_echo_correlation_eq}
   -\frac{\ell^2}{2}\Delta C_D^{(4)}(\bm{r})=-ik_0\ell\chi\bra I(\bm{r})\ket^2.
\end{equation}
This equation is the main result from the analytical theory. It takes a similar form as \eq{eq:diffusion} obtained above for the diffuse intensity.
However, in sharp contrast to \eq{eq:diffusion}, the source term in \eq{eq:diffuse_echo_correlation_eq} has significant
values at any depth in the medium, which entails a twofold consequence. First, since the correlation buildup is not
localized near the slab input, the bulk approximation made for the diffusion equation is justified. There is not need to
try to extrapolate this equation outside its region of validity. Second, the continuous feeding of $C_D^{(4)}(\bm{r})$ by
$\bra I(\bm{r})\ket^2$ strongly contributes to enhance the correlation, all along the progression through the medium.   

Provided \eq{eq:diffuse_echo_correlation_eq} is complemented with the boundary conditions [see \eq{eq:boundary_cond}]
previously used to solve \eq{eq:diffusion}, the solution $C_D^{(4)}(z)$ is also given by
\eq{eq:diffusion_eq_solution}, where $F(z)$ now represents any solution of
\begin{equation}
   F''(z)=-2i\frac{k_0}{\ell}\chi \bra I(z)\ket^2.
\end{equation}

We may compare the correlation in a disordered medium with the corresponding quantity in an homogeneous slab. The latter reads as    
\begin{multline}\label{eq:correlation_echo_hom}
   C^{(4)}_{\text{hom}}(z)=E_{\text{hom}}(z)E^{(4)*}_{\text{hom}}(z)
\\
      =\frac{\chi I_0^2}{4}\left\{-2ik_0z-1+\exp[-2ik_0(L-z)]\right\}
\end{multline}
which, for depths larger than $\lambda=2\pi/k_0$, becomes 
\begin{equation}
   C^{(4)}_{\text{hom}}(z)=\frac{-ik_0\chi I_0^2 z}{2}=i\sqrt{I_0I_{\text{hom}}^{(4)}(z)}.
\end{equation} 

\begin{figure}[!htb]
   \centering
   \psfrag{z}[c][t]{$z/\ell$}
   \psfrag{AAAAAA2}[l][l]{$\left|C^{(4)}(z)\right|/\left|C^{(4)}_{\text{hom}}(L)\right|$ numerical}
   \psfrag{BBBBBB2}[l][l]{$\left|C^{(4)}(z)\right|/\left|C^{(4)}_{\text{hom}}(L)\right|$ analytical}
   \includegraphics[width=0.9\linewidth]{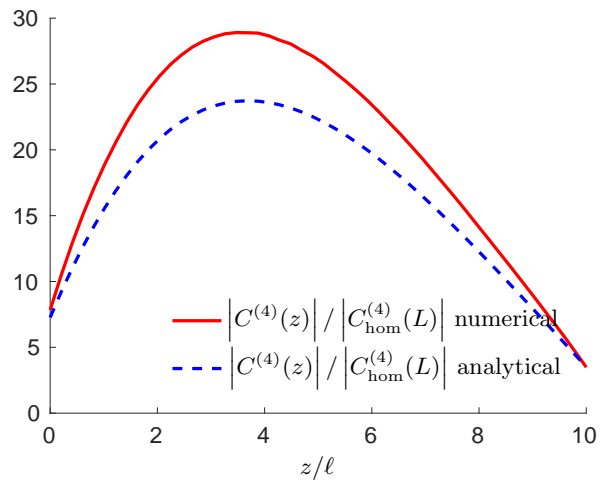}
   \caption{(Color online) Correlation of the echo signal with one of the driving fields $C^{(4)}=\bra
   EE^{(4)*}\ket$ as a function of the depth $z/\ell$ inside the slab (analytical solution:blue dashed
   line; numerical computation:solid line); $b=10$ and $k_0\ell=40$. The correlation is normalized to the
   corresponding quantity at the exit of a homogeneous slab.}
   \label{fig:echo_correlation}
\end{figure}

According to \figs{fig:echo_correlation} where we have displayed the variations of
$\left|C_D^{(4)}(z)\right|/\left|C^{(4)}_{\text{hom}}(L)\right|$ with $z$, the strength of $\left|C_D^{(4)}(z)\right|$
largely exceeds that of $\left|C^{(4)}_{\text{hom}}(L)\right|$ at any depth.

Normalization with $\sqrt{\bra I\ket\bra I^{(4)}\ket}$, where $\bra I^{(4)}\ket$ stands for the echo average intensity,
helps to reveal the correlation strength. Indeed, as a consequence of the Cauchy Schwarz inequality, the variation range
of $\left|C_D^{(4)}(z)\right|/\sqrt{\bra I(z)\ket\bra I^{(4)}(z)\ket}$ is limited to interval $[0,1]$, where the upper
bound is reached when the echo is fully correlated with the reference field. In order to obtain the normalized
correlation, we calculate the echo average intensity in the next section.

\subsection{Average intensity of the echo and normalized correlation}

\begin{figure}[!htb]
   \centering
   \psfrag{z}[c][t]{$z/\ell$}
   \psfrag{AAAAAA}[l][l]{$\bra I^{(4)}\ket/I_{\text{hom}}^{(4)}(L)$ numerical}
   \psfrag{BBBBBB}[l][l]{$\bra I^{(4)}\ket/I_{\text{hom}}^{(4)}(L)$ analytical}
   \includegraphics[width=0.9\linewidth]{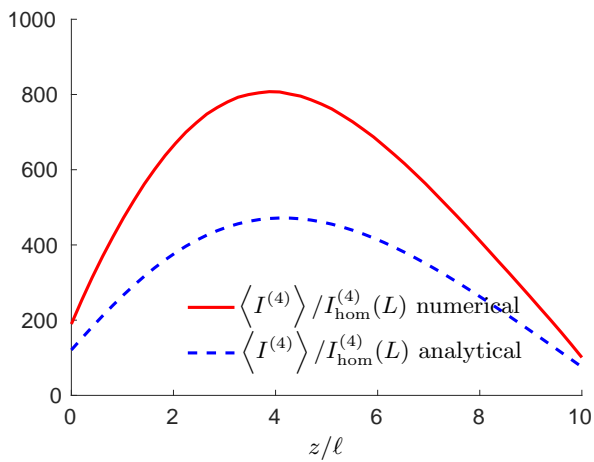}
   \caption{(Color online) Echo intensity $\bra I^{(4)}\ket=\bra\left|E^{(4)}\right|^2\ket$ as a function of the depth
   $z/\ell$ inside the slab (numerical computation: red solid line; analytical solution: blue dashed line). $b=10$ and
   $k_0\ell=40$.}
   \label{fig:average_echo_intensity}
\end{figure}

As for the incoming intensity and the calculation of the correlation function, we expand the echo average intensity into
a ballistic and a diffuse part as follows:
\begin{equation}\label{eq:average_echo_intensity}
   \bra I^{(4)}(\bm{r})\ket=I_B^{(4)}(\bm{r})+I_D^{(4)}(\bm{r}).
\end{equation}
The ballistic part is given by \eq{eq:ballistic_echo_intensity} and the diffuse part reads diagrammatically as
\begin{multline}
   I_D^{(4)}(\bm{r}) = 
\\
   \begin{dddddiag}{54}
      \ggmoy{0}{6}{-3}
      \ggmoy{6}{12}{-3}
      \ggmoy{12}{18}{-3}
      \ggmoy{18}{24}{-3}
      \ggmoy{24}{30}{-3}
      \ggmoy{30}{36}{-3}
      \ggmoy{36}{48}{-3}
      \eemoy{48}{54}{-3}
      \ggmoy{0}{6}{3}
      \ggmoy{6}{12}{3}
      \ggmoy{12}{24}{3}
      \ggmoy{24}{30}{3}
      \ggmoy{30}{36}{3}
      \ggmoy{36}{42}{3}
      \ggmoy{42}{48}{3}
      \eemoy{48}{54}{3}
      \iidentique{6}{-3}{6}{3}
      \iidentique{12}{-3}{12}{3}
      \iidentique{24}{-3}{24}{3}
      \iidentique{30}{-3}{30}{3}
      \iidentique{42}{-3}{42}{3}
      \iidentique{48}{-3}{48}{3}
      \pparticule{6}{-3}
      \pparticule{12}{-3}
      \pparticule{24}{-3}
      \pparticule{30}{-3}
      \pparticule{42}{-3}
      \pparticule{48}{-3}
      \pparticule{6}{3}
      \pparticule{12}{3}
      \pparticule{24}{3}
      \pparticule{30}{3}
      \pparticule{42}{3}
      \pparticule{48}{3}
      \gggmoy{36}{42}{3}{9}
      \gggmoy{36}{42}{3}{15}
      \ggmoy{42}{48}{9}
      \eemoy{48}{54}{9}
      \ggmoy{42}{48}{15}
      \eemoy{48}{54}{15}
      \iidentique{42}{15}{42}{9}
      \iidentique{48}{15}{48}{9}
      \pparticule{42}{9}
      \pparticule{48}{9}
      \pparticule{42}{15}
      \pparticule{48}{15}
      \gggmoy{18}{24}{-3}{-9}
      \gggmoy{18}{24}{-3}{-15}
      \ggmoy{24}{30}{-9}
      \ggmoy{30}{36}{-9}
      \ggmoy{36}{42}{-9}
      \ggmoy{42}{48}{-9}
      \eemoy{48}{54}{-9}
      \ggmoy{24}{30}{-15}
      \ggmoy{30}{36}{-15}
      \ggmoy{36}{42}{-15}
      \ggmoy{32}{48}{-15}
      \eemoy{48}{54}{-15}
      \iidentique{24}{-15}{24}{-9}
      \iidentique{30}{-15}{30}{-9}
      \iidentique{36}{-15}{36}{-9}
      \iidentique{42}{-15}{42}{-9}
      \iidentique{48}{-15}{48}{-9}
      \pparticule{24}{-9}
      \pparticule{30}{-9}
      \pparticule{36}{-9}
      \pparticule{42}{-9}
      \pparticule{48}{-9}
      \pparticule{24}{-15}
      \pparticule{30}{-15}
      \pparticule{36}{-15}
      \pparticule{42}{-15}
      \pparticule{48}{-15}
      \nnonlineaire{18}{-3}
      \nnonlineaire{36}{3}
   \end{dddddiag}
\\
   +\ldots
\end{multline}
We follow the same procedure as for the echo correlation function $C_D^{(4)}$. According to the diagram, the source term
for the echo diffuse intensity reads as the correlation multiplied by the average intensity. This leads to the following
diffusion equation governing the evolution of $I_D^{(4)}$:
\begin{equation}\label{eq:diffuse_echo_intensity}
   -\frac{\ell^2}{2}\Delta I_D^{(4)}(\bm{r})
      =2k_0\ell\re\left[i\chi^* C^{(4)}(\bm{r})\right]\bra I(\bm{r})\ket.
\end{equation}
Again, the source term in this diffusion equation is delocalized over the whole sample, thus leading to a different
behavior for $I_D^{(4)}$ compared to $I_D$ (or for $\bra I^{(4)}\ket$ compared to $\bra I\ket$) even if both quantities
have significant values for all depths inside the slab, as illustrated in \figs{fig:average_echo_intensity} and
\ref{fig:ballistic_diffuse}. 

We note that $\bra I^{(4)}\ket$ is much larger than  $I^{(4)}_{\text{hom}}$. Two arguments can be put forward as an
explanation. (1) In a random walk picture, the paths followed by photons inside the disordered medium can be much longer
than the slab thickness. Indeed, the average path length is of the order of $\bra s\ket=2L^2/\ell$ in a thick ($b\gg 1$)
and dilute ($k_0\ell\gg 1$) scattering medium while it is $s_{\text{hom}}=L$ for an homogeneous slab. (2) In a
disordered medium, the driving fields have larger values than in an homogeneous material thanks to light confinement
by scattering. This is visible in \fig{fig:ballistic_diffuse} where a maximum average intensity on the order of $2.5I_0$
is reached.

\begin{figure}[!htb]
   \centering
   \psfrag{z}[c][t]{$z/\ell$}
   \psfrag{AAAAAA1}[l][l]{$\left|C^{(4)}\right|/\sqrt{\bra I\ket\bra I^{(4)}\ket}$ numerical}
   \psfrag{BBBBBB1}[l][l]{$\left|C^{(4)}\right|/\sqrt{\bra I\ket\bra I^{(4)}\ket}$ analytical}
   \includegraphics[width=0.9\linewidth]{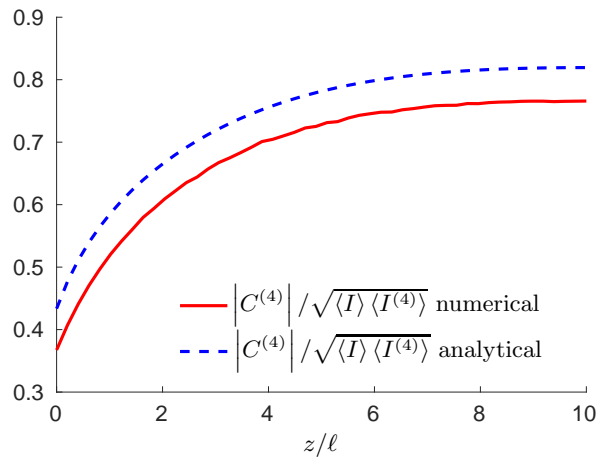}
   \caption{(Color online) Normalized correlation $\left|C^{(4)}(z)\right|/\sqrt{\bra I(z)\ket\bra I^{(4)}(z)\ket}$
   inside the slab. The analytical result, based on the diagrammatic approach, and the numerical computation are both
   displayed as a function of the normalized depth $z/\ell$ (blue dashed line and red solid line respectively), with
   $b=10$ and $k_0\ell=40$.}
   \label{fig:echo_correlation_norm}
\end{figure}

Finally, \fig{fig:echo_correlation_norm} shows the dramatic increase of the normalized correlation
$\left|C^{(4)}(z)\right|/\sqrt{\bra I(z)\ket\bra I^{(4)}(z)\ket}$ with the penetration depth, up to $\approx 0.8$ at the
slab exit.

\section{Numerical results}\label{sec:numerics}

The analytical expressions are expected to be consistent with the statistically averaged solutions of the coupled wave
equations (see Sec.~\ref{subsec:wave_equations}). We must resort to numerical computation to obtain these solutions, in
order to validate the analytical approach. 

Solving \eqs{eq:coupled_dipoles} represents the most challenging task. Indeed, to solve this set of $N_s$ equations, one
has to inverse a large $N_s\times N_s$ matrix, the actual size of which is imposed by the large depth ($L\gg \ell$) and
slab geometry (infinite transverse extension) assumptions. In order to minimize $N_s$, we reduce the slab transverse
dimension to $D=4L$, expected to be a good trade-off, limiting finite-size effects while maintaining a reasonable computing
time. To satisfy the diffusive-regime, large-depth, condition we set $L/\ell=10$. Hence $N_s=DL\rho_s=400\ell^2\rho_s$.
According to \eq{eq:ell}, for a given value of $\ell$, the scatterer density $\rho_s$ is minimized when $\im\alpha$ is
maximized. As already pointed out in Sec.~\ref{subsec:wave_equations} (see \eq{eq:alpha_max}), the maximum value of
$\im\alpha$, compatible with energy conservation, is $4/k_0^2$, which leads to $\ell^2\rho_s=k_0\ell/4$. Finally, to satisfy
the dilute medium condition $k_0\ell\gg 1$, we set $k_0\ell=40$, which leads to $N_s=4000$. 

Since the echo signal \eqs{eq:coupled_dipoles_echo} and the driving field \eqs{eq:coupled_dipoles} only differ from each
other through the source term, they are both solved by the same inverse matrix. In the echo signal equations we have to
discretize the integral of the source term. As we perform statistics (\ie computation of average fields, intensities and
correlations), we have chosen to treat the active region as a collection of $N_a$ randomly placed active atoms at
positions $\bm{\uprho}_j$. This leads to

\begin{multline}
   k_0^2\chi\int G_0(\bm{r}-\bm{r}')E^{(1)*}(\bm{r}')E^{(2)}(\bm{r}')E^{(3)}(\bm{r}')\ud\bm{r}'
\\
      \sim k_0^2\chi S_a\sum_{j=1}^{N_a} G_0(\bm{r}-\bm{\uprho}_j)E^{(1)*}(\bm{\uprho}_j)E^{(2)}(\bm{\uprho}_j)E^{(3)}(\bm{\uprho}_j)
\end{multline}
where $S_a=LD/N_a$ is the surface of one active atom. The statistical observables are not sensitive
to the number of active regions $N_a$, even if very small (\ie continuum not reached).  In practice, we have used
$N_a=4000$. 

Having obtained the expression of the exciting echo field on each scatterer, we use \eq{eq:coupled_dipoles_echo_field}
to compute the echo field at any position in or outside the system. 

Repeating the same procedure for a large set of randomly drawn configurations, we are in position to evaluate
statistical quantities such as the average echo field, the correlation of the echo with a driving field, and the average
echo intensity. 

In slab geometry, under plane-wave illumination at normal incidence, statistical quantities are invariant by translation
along the transverse direction $x$. To spare computation time, while taking care of finite-size effects, we combine
average over $N_{\text{conf}}=200000$ configurations with limited range integration over $x$.

\subsection{Average field of the echo}

As observed in \fig{fig:ballistic_echo_intensity}, the analytical and the numerical approaches consistently describe the
echo ballistic intensity variation with the depth inside the slab, although the analytical expression is derived under
the diffusion approximation, valid only at large depths (\ie $z\gg\ell$).

\subsection{Correlation of photon echoes with the driving fields}

\Fig{fig:echo_correlation} gives the evolution of the correlation of the echo field with one of the driving fields as a
function of the depth inside the slab. Again a good agreement is clearly obtained between the analytical calculation and
the numerical model. Nevertheless, the analytical result is not fully quantitative. Two potential effects have been
identified to explain this discrepancy. First the validity of the diffusion approximation can be questioned. In the
linear regime, the agreement between the diffusion equation theory and the coupled-dipole simulation is very good as
shown in \fig{fig:ballistic_diffuse} even for small depths. However when non-linearities are present, there is
potentially an accumulation of errors because of the recursion in the diffusion model provided by \eqs{eq:diffusion} and
\eqref{eq:diffuse_echo_correlation_eq}. For small and intermediate depths, the Radiative Transfer Equation (RTE) could
be a good candidate for a refined model valid at all depth~\cite{CHANDRASEKHAR-1950}. However the main drawback is that
analytical results do not exist for the RTE in a slab geometry. Second and potentially more important, the signal is
very sensitive to the boundaries in the presence of non-linear effects. This has been checked numerically by changing
the transverse size $D$ of the pseudo-slab geometry and the results show that converged results are hard to obtain.

\subsection{Average intensity of the echo}

Regarding the average intensity of the echo signal, the numerical results are presented in
\fig{fig:average_echo_intensity}. Although qualitative agreement is preserved (confirming that the analytical theory
captures the main physical mechanisms), a larger discrepancy is found between the theory and the simulations than for
the correlation. The reasons are the same: finite transverse size effects in presence of non-linearity and validity of
the diffusion approximation.

\section{Conclusion}\label{sec:conclusion}

We have presented a theoretical study of photon echo generation in disordered scattering media. Developed in terms of
Feynman-Dyson diagrams, the multiple scattering statistical approach has been validated by \textit{ab initio} numerical
simulations.

According to previous experiments~\cite{beaudoux2011}, the driving fields and the echo beam stay strongly correlated as
they propagate through the disordered medium. The theory has confirmed this paradoxical feature, and provided some
physical insight. In the buildup of any two-field observable, such as diffuse intensity or diffuse correlation, the same
dominant diagrams emerge: those that make both fields follow a common path through the disordered medium. This single
propagation scheme explains the similar size of those different quantities, and the large size of the normalized
correlation.

Another noticeable result is the strong enhancement of the echo by the disordered medium, in comparison with echo
emission in the corresponding homogeneous material with the same concentration of active atoms. This might open the way
to applications in energy conversion.
    
The present work has been confined to signal investigation inside the disordered material. To be consistent with
experimental conditions, we should consider signal collection outside the material, on a large aperture detector. This
issue is deferred to a future work. Encouraged by the present promising results, we also plan to refine the analysis in
such directions as that of the RTE, with the help of Monte Carlo simulations.

\section*{Acknowledgments}

We thank Philippe Goldner for the stimulated discussions that initiated this work. We are also grateful to Thierry
Chaneli\`ere for helpful comments and advices. This research is supported by the French national grant RAMACO
no.~ANR-12-BS08-0015-02 and by LABEX WIFI (Laboratory of Excellence within the French Program ``Investments for the
Future'') under references ANR-10-LABX-24 and ANR-10-IDEX-0001-02 PSL*.

\end{document}